    \documentstyle[epsf]{article} \catcode`\@=11
    \def\R{{\rm I\!R}}
    \def\section{\@startsection{section}{1}{\z@}%
    {-3.5ex plus -1ex minus -.5ex}{1.5ex plus.3ex}{\bf }}
    \def\subsection{\@startsection{subsection}{1}{\z@}%
    {-3.5ex plus-1ex minus-.5ex}{1.5ex plus.3ex}{\bf }} 
    
\begin{document}
    \hfill\parbox{4.77cm}{\Large\centering Annalen\\der
    Physik\\[-.2\baselineskip] {\small \underline{\copyright\ Johann
    Ambrosius Barth 1998}}} \vspace{.75cm}\newline{\Large\bf
Numerical tests of conjectures of conformal field theory for
three-dimensional systems
    }\vspace{.4cm}\newline{\bf   
Martin Weigel and Wolfhard Janke
    }\vspace{.4cm}\newline\small
Institut f\"{u}r Theoretische Physik, Universit\"{a}t Leipzig,
D-04109 Leipzig, Germany, and\\
  Institut f\"{u}r Physik, Johannes Gutenberg-Universit\"{a}t Mainz,
D-55099 Mainz, Germany\\
  {\tt Martin.Weigel@itp.uni-leipzig.de, Wolfhard.Janke@itp.uni-leipzig.de}
    \vspace{.2cm}\newline 
Received 6 October 1998, accepted 8 October 1998 by U. Eckern
    \vspace{.4cm}\newline\begin{minipage}[h]{\textwidth}\baselineskip=10pt
    {\bf  Abstract.}
The concept of conformal field theory provides a general classification of
statistical systems on two-dimensional
geometries at the point of a continuous phase transition.
Considering the finite-size scaling
of certain special observables, one thus obtains not only
the critical exponents but even the corresponding {\em amplitudes} of the
divergences analytically.
A first numerical analysis brought up the question whether analogous
results can be obtained for those systems on three-dimensional manifolds.

Using Monte Carlo simulations based on the Wolff single-cluster
update algorithm we investigate the scaling properties of
O($n$) symmetric classical spin models on a
{\em three-dimensional}, hyper-cylindrical geometry with a toroidal
cross-section considering both periodic and antiperiodic boundary conditions.
Studying the correlation lengths of
the Ising, the XY, and the Heisenberg
model, we find strong evidence for a scaling relation
analogous to the two-dimensional case, but in contrast here for the systems
with {\em antiperiodic} boundary conditions.
    \end{minipage}\vspace{.4cm} \newline {\bf  Keywords:}
Spin models; Finite-size scaling; Universal amplitudes;
Conformal field theory
    \newline\vspace{.2cm} \normalsize
\section{Introduction}
Statistical mechanical systems at a critical point are essentially characterized
by a loss of length scales: as the correlation length diverges, the system becomes
a self-similar random fractal. Augmenting this symmetry with translational and
rotational invariance in the continuum limit establishes the so-called {\em conformal}
invariance of the system. As the 2D conformal group is of infinite dimension, exploiting
this feature allows a complete classification of models of statistical mechanics
according to their operator content in two dimensions \cite{CardyBuch,HenkelBuch}. 
This includes a special class of formally model independent relations, like finite-size
scaling (FSS) laws;
in particular, for the 2D strip geometry $S^1\times\R$ with periodic boundary conditions
Cardy \cite{Cardy84a} has shown, that for any
primary, i.e. conformally covariant, operator of a model showing critical behavior
the corresponding correlation length scales as:
\begin{equation}
  \xi_i=\frac{A}{x_i}L,
\label{amplit}
\end{equation} 
where $L$ denotes the circumference of the cylinder, $x_i$ is the scaling dimension
of the considered operator, a combination of the classical critical exponents,
and $A=1/2\pi$ in the 2D case.

For 3D systems, however, the situation is different. First, the concept of a primary
operator becomes at least problematic (it might possibly be established in terms of the
operator product expansion (OPE) \cite{CardyPrivate}). Secondly, numerically
feasible and more widely applicable geometries like that of a column
$S^1\times S^1\times\R$ are not conformally related to flat spaces like in the
above mentioned 2D case, which is a essential ingredient of the derivation of relation
(\ref{amplit}). A transfer matrix calculation by Henkel \cite{Henkel86,Henkel87} for
the $S=\frac{1}{2}$ Ising model on the column geometry gave for the ratio of the
correlation lengths of the magnetization and energy densities the values
$\xi_\sigma/\xi_\epsilon=3.62(7)$ and $2.76(4)$ for periodic and antiperiodic
boundary conditions, respectively. Comparing this with the ratio of scaling
dimensions of $x_\epsilon/x_\sigma=2.7326(16)$ \cite{diplom} results in the astonishing,
theoretically obscure, conjecture that a relation of the form (\ref{amplit}) can
be re-established in the 3D case, when applying {\em antiperiodic} (apbc) instead of
periodic (pbc) boundary conditions along the torus directions, which later on could
be affirmed by a Metropolis Monte Carlo (MC) simulation by Weston \cite{Weston}.

If this result could be established analytically, it would be one of the rare
rigorous statements for non-trivial 3D systems. As simulational data were available
up to this point only for the single special case of the Ising model, we thought it
rewarding to analyze some further models, so possibly establishing this conjecture
at an empirical level, which should constitute a motivation and basis for further
analytical studies.

\section{Models}
In generalization to the Ising case we restrict ourselves to the class of O($n$)
symmetric classical spin models with Hamiltonian
\begin{equation}
\label{Hamilton}        
{\cal H} = -J \sum_{<ij>} {\bf s}_i\cdot{\bf s}_j,\;\;{\bf s}_i \in S^{n-1},
\end{equation}
assuming nearest-neighbor, ferromagnetic ($J>0$)
interactions. The simulations were done for a discrete sc lattice with
dimensions $(L_x,L_y,L_z)$, choosing $L_x=L_y$ and $\xi/L_z\ll 1$, therewith
approximately modelling the column geometry $S^1\times S^1\times\R$.

\begin{figure}[tb]
\begin{picture}(250,200)
    \put(0, 0){\includegraphics{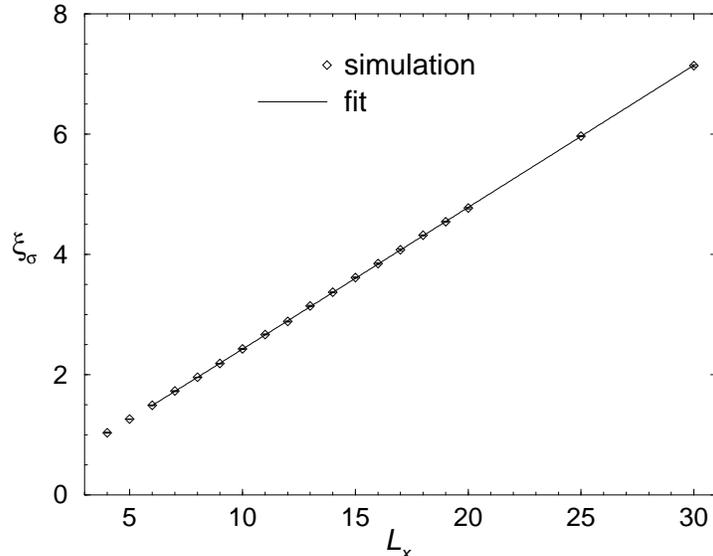}}
  \end{picture}
\caption{
FSS plot for the spin correlation length $\xi_\sigma(L_x)$ of the
3D Ising model with antiperiodic boundary conditions. The solid line represents
a least-square fit according to Eq.\ (\ref{fitform}).}
\label{fig1}
\end{figure}

\section{Simulation and data analysis}
The MC simulations were done using the Wolff single-cluster updating scheme
\cite{Wolff89}, as it is known to be more efficient than the Swendsen-Wang
\cite{Swendsen} update in three dimensions \cite{WJChem}. In order to be able
to perform simulations for both, periodic and antiperiodic boundary conditions,
the Wolff update had to be adapted to the latter case: this was achieved by
exploiting the fact that antiperiodic bc are equivalent to the insertion of a
seam of anti-ferromagnetic bonds  along the boundary in the case of
nearest-neighbor interactions.

To test for a relation according to (\ref{amplit}) we had to measure at least
two different correlation lengths of the systems under consideration. Following
Henkel and in an analogy to the 2D Ising case, where the only
non-trivial primary operators are the densities of magnetization and energy, we
recorded the correlation functions of these two operators:
\begin{equation}
\begin{array}{rcl}
  G_{\sigma}^c({\bf x}_1,{\bf x}_2) & = & \langle{\bf s}({\bf x}_1)\cdot{\bf
  s}({\bf x}_2)\rangle-\langle{\bf s}\rangle\langle{\bf s}\rangle, \\
  G_{\epsilon}^c({\bf x}_1,{\bf x}_2) & = & \langle\epsilon({\bf
  x}_1)\,\epsilon({\bf
  x}_2)\rangle-\langle\epsilon\rangle\langle\epsilon\rangle. \\
\end{array}
\label{conncorr}
\end{equation}
Variance-reduction of the estimators for these observables can be achieved in a
first step by the trivial average over values with
$({\bf x}_1-{\bf x}_2)\parallel \hat{e}_z$ and
$i\equiv|{\bf x}_1-{\bf x}_2|=\mbox{const}$; they can be further improved by
applying a zero momentum mode projection, i.e., by summing up the values for
the densities in the layers $z=\mbox{const}$ before correlating them \cite{WJ93}.

\begin{figure}[tb]
\begin{picture}(250,200)
    \put(0, 0){\includegraphics{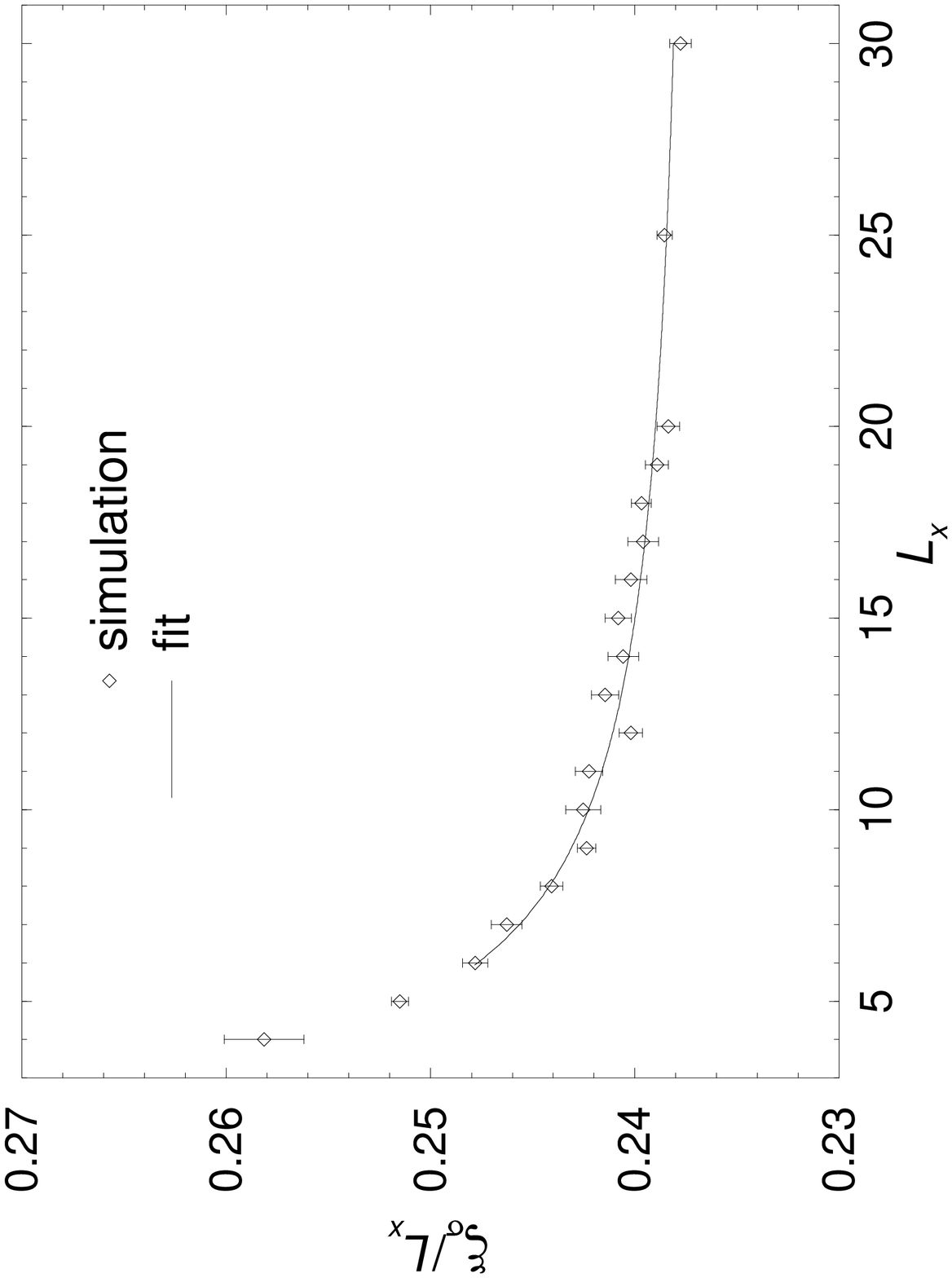}}
  \end{picture}
\caption{
Scaling of the amplitudes $\xi_\sigma/L_x$ of the 3D Ising model with
antiperiodic boundary conditions.}
\label{fig2}
\end{figure}

\begin{table}[tb]
\caption{Finite-size scaling amplitudes of the correlation lengths of the Ising, XY, and
Heisenberg models on the $T^2\times\R$ geometry.
\label{tab1}}
\begin{center}
\begin{tabular}{||clll||} \hline
model & & \multicolumn{1}{c}{pbc} & \multicolumn{1}{c||}{apbc} \\ \hline
      & $A_\sigma$    & 0.8183(32) & 0.23694(80) \\
      & $A_\epsilon$  & 0.2232(16)  & 0.08661(31) \\
\raisebox{1ex}[-1ex]{Ising} & $A_\sigma/A_\epsilon$ & 3.666(30) & 2.736(13) \\ 
      & $x_\epsilon/x_\sigma$ &  \multicolumn{2}{c||}{2.7326(16)} \\ \hline
      & $A_\sigma$    & 0.75409(59) & 0.24113(57) \\
      & $A_\epsilon$  & 0.1899(15)  & 0.0823(13) \\ 
\raisebox{1ex}[-1ex]{XY} & $A_\sigma/A_\epsilon$ & 3.971(32) & 2.930(47) \\ 
      & $x_\epsilon/x_\sigma$ & \multicolumn{2}{c||}{2.923(7)} \\ \hline
      & $A_\sigma$    & 0.72068(34) & 0.24462(51) \\
      & $A_\epsilon$  & 0.16966(36)  & 0.0793(20) \\
\raisebox{1ex}[-1ex]{Heisenberg} & $A_\sigma/A_\epsilon$ & 4.2478(92) & 3.085(78) \\ 
      & $x_\epsilon/x_\sigma$ &  \multicolumn{2}{c||}{3.091(8)} \\ \hline
\end{tabular}
\end{center}
\end{table}

For extracting the correlation lengths from (\ref{conncorr}) one can cancel out
deviations that arise from inaccuracies in the determination of the disconnected
parts of the correlation functions and eliminate the need for a correct
normalization of the estimates by considering the following set of estimators:
\begin{equation}
\hat{\xi}_i=\Delta{\left[\ln\frac{\hat{G}^{c,\parallel}(i)-\hat{G}^{c,\parallel}(i-\Delta)}
{\hat{G}^{c,\parallel}(i+\Delta)-\hat{G}^{c,\parallel}(i)}\right]}^{-1},
\label{diffmethoddelta}
\end{equation}
where $\Delta\ge 1$ should usually be chosen so that a constant drop of $G(i)$
between the pairs of considered points is guaranteed. Variances and
cross-correlations of the $\hat{\xi}(i)$ were estimated using a combined binning
\cite{Flyvbjerg} and jackknifing \cite{Efron} technique. In a process of
statistical optimization, resulting in the leaving out of the corrupt estimates
$\hat{\xi}(i)$ for distances in the regions $i<\Delta$ and $i>L_z/2$, one ends
up with a final value for the correlation lengths $\xi_\sigma$ and
$\xi_\epsilon$ of the considered system.

All simulations were done at inverse temperatures, which were either highly
precise single estimates of the inverse critical temperature of the bulk model
or weighted means of several such estimates \cite{cond,ours}, the influence of
uncertainties in these values being checked via a temperature reweighting
technique.

\section{Results}
The cumulated estimates for the correlation lengths $\hat{\xi}(L_x)$ for the
different system sizes exhibit an almost perfect linear scaling behavior as shown
in Fig. \ref{fig1} for the Ising model and antiperiodic bc. For all models we
analyzed system sizes between $L_x=4$ and $30$ and volumes up to about $3\cdot
10^5$ spins. The plot of the amplitudes $\hat{\xi}/L_x$ in Fig. \ref{fig2},
however, reveals a clear resolution of corrections to scaling, however.
Therefore fits to a law including corrections of the form
\begin{equation}
\xi(L_x)=AL_x+BL_x^{\alpha}
\label{fitform}
\end{equation}
were done to arrive at estimates for the leading order scaling amplitudes
$A_\sigma$ and $A_\epsilon$.

The final results for these leading amplitudes are summarized in
Table \ref{tab1}. The values for the scaling dimensions shown for comparison
are once again weighted literature means \cite{diplom}.

\section{Conclusions}
The clear conclusion of these
results for all three models under consideration is that, while for the generic
case of periodic bc the ratios of the amplitudes and the scaling dimensions
differ by at least about thirty sigma, in the case of {\em antiperiodic} bc both
ratios agree to a very high level of precision, thus giving the conjecture of a
law equivalent to (\ref{amplit}) for 3D models enough backing to think
seriously about a theoretical justification. So, for the $n=1,2,3$ representatives
of the class of O($n$) spin models
one can state that the finite-size scaling amplitude ratios of the
the correlation lengths of the magnetization and energy densities
is determined by the corresponding scaling dimensions
and thus universal; in connection with additional results for the case
$n=10$ \cite{ours}
and an analytical results for the spherical model \cite{Henkel88}, it seems
reasonable to assume that such a relation holds for the whole class of O($n$)
spin models. Note, however, that the amplitude $A$ in Eq. (\ref{amplit}), which
was $1/2\pi=\mbox{const}$ in the 2D case, does now depend on the model under
consideration, i.e., the dimension $n$ of the order parameter \cite{cond}.
    \vspace{0.6cm}\newline{\small 
We thank K. Binder for his constant and generous support. We are grateful to
J. Cardy and M. Henkel for helpful discussions on the theoretical
background. W.J. gratefully acknowledges support from the Deutsche
Forschungsgemeinschaft through a Heisenberg Fellowship.
    }
    \end{document}